\documentclass[10pt,journal,compsoc]{IEEEtran}
\usepackage{amssymb}
\usepackage{graphicx}
\usepackage{balance}
\usepackage{array,colortbl,multirow,multicol,booktabs,ctable}
\usepackage{url}
\usepackage[english]{babel}
\usepackage[misc]{ifsym}
\usepackage{bbding}
\usepackage{balance}
\usepackage{algorithmic}
\usepackage[Algorithm,ruled]{algorithm}
\usepackage{listings}
\usepackage{diagbox}
\usepackage{hyperref}
\usepackage{xcolor}
\hypersetup{
    colorlinks,
    linkcolor={blue!50!black},
    citecolor={blue!50!black},
    urlcolor={blue!80!black}
}

\makeatletter
  \newcommand\tinyv{\@setfontsize\tinyv{8.2pt}{6}}
  \newcommand\tinycode{\@setfontsize\tinycode{8.2pt}{6}}
\makeatother

\begin{document}

\title{JNI Global References Are Still Vulnerable: Attacks and Defenses}

\author{Yi He, Yuan Zhou, Yajin Zhou, Qi~Li,~\IEEEmembership{Senior~Member,~IEEE,} Kun Sun, \\ Yacong~Gu, Yong Jiang
\thanks{Yi He and Yuan Zhou contributed equally to this manuscript.}
\thanks{Yi He and Qi Li are with Institute for Network Sciences and Cyberspace, Tsinghua University, Beijing, China 100084, and Beijing National Research Center for Information Science and Technology (BNRist), Beijing, China 100084, e-mail:\{heyi14@mails., qli01@\}tsinghua.edu.cn.}

\thanks{Yuan Zhou and Yong Jiang are with Tsinghua-Berkeley Shenzhen institute, Tsinghua  University,  Shenzhen,  China  518055, e-mail:\{yuan-zho17@mails., jiangy@sz.\}tsinghua.edu.cn.}

\thanks{Yajin Zhou is with School of Cyber Space and Technology and College of Computer Science and Technology, Zhejiang University, China 310058, email: yajin\_zhou@zju.edu.cn.}
\thanks{Kun Sun is with George Mason University at Department of Information Sciences and Technology, USA,  Fairfax, VA 22030-4422, email: ksun3@gmu.edu}
\thanks{Yacong Gu is with Institute of Software, Chinese Academy of Sciences,  China 100190, e-mail: \{guyangcong\}@tca.iscas.ac.cn.}
\thanks{Corresponding authors: Yajin Zhou and Qi Li.}}

\IEEEtitleabstractindextext{
\begin{abstract}

System services and resources in Android are accessed through IPC based mechanisms. Previous research has demonstrated that they are vulnerable to the denial-of-service attack (DoS attack). For instance, the JNI global reference (JGR), which is widely used by system services, can be exhausted to cause the system reboot (hence the name JGRE attack). Even though the Android team tries to fix the problem by enforcing  security checks, we find that it is still possible to construct a JGR exhaustion DoS attack in the latest Android system.

In this paper, we propose a new JGR exhaustion DoS attack, which is effective in different Android versions, including the \textit{latest one (i.e., Android 10)}. Specifically, we developed JGREAnalyzer, a tool that can systematically detect JGR vulnerable services APIs via a call graph analysis and a forwarding reachability analysis. We applied this tool to different Android versions and found multiple vulnerabilities. In particular, among 148 system services in Android 10, 12 of them have 21 vulnerabilities. Among them, 9 can be successfully exploited without any permissions. We further analyze the root cause of the vulnerabilities and propose a new defense to mitigate the JGRE attack by restricting resource consumption via global reference counting.

\end{abstract}
}

\maketitle

\section{Introduction}

Android is the most widely used mobile operating system. Due to its
popularity, it becomes the primary target of different attacks.
In our previous work,
a special attack called Java Global Reference Exhaustion
(JGRE) attack~\cite{Gu2017JGRE} was reported in Android 6. 
The malicious app continuously invokes vulnerable
system APIs so that it will create a large number of Java Global Reference (a special
Java object) in system services. Since it consumes lots of
memory, it eventually will cause the system service crash.

Moreover, if the attack is targeting the core services of the Android system,
it could cause the system to reboot. For instance, if we successfully performed the JGRE attack to the \textit{ServiceManager}, the Android phone will reboot. Note that, this type of attack could be easily launched without any
requirement of permissions, or even could be remotely launched via the web application frameworks such as Cordova \cite{Cordova}.
Due to the potential consequences that could be caused and the easy manner to launch the attack, countermeasures to this attack should be proposed.

Google has realized this problem and applied multiple patches since Android 6
to mitigate this problem. However, we find that the mitigation is in an
ad-hoc manner, and thus can be bypassed. For instance, 
Google fixed 32 vulnerable interfaces in Android 8.1 by removing
vulnerable interfaces or elevating permission level to access these
interfaces. However, this fix can still be bypassed, by leveraging
the un-patched vulnerable interfaces. In the latest Android version
(Android 10), Google limits the binder proxy in each process by setting a threshold in the create method of binder proxy. It kills the process once it creates over 6000 binder proxies. Nevertheless, we have found a new JGRE attack that can bypass this
countermeasure.

The problem of the current defense used by Google is that it lacks a
systematic way to assess this vulnerability. Specifically, it does not
have an automatic way to find all potential attack interfaces.
To solve this problem, we propose and implement a static analysis tool
called JGREAnalyzer, which analyzes the implementation of system services and
finds all potential JGRE vulnerabilities. Specifically, it performs
forward reachability analysis from the Java code to the native code to find
potentially vulnerable code paths. Then it dynamically verifies
these paths by automatically generating inputs to trigger the vulnerabilities. 
By doing so, JGREAnalyzer can identify vulnerable IPC interfaces that could
be exploited by malicious apps.

We applied JGREAnalyzer to various Android versions from Android 6 to Android
10, and found multiple vulnerabilities. In particular, in Android 10 with
the latest defense mechanism enforced, our tool found 12 system services that are vulnerable. Among them, 9 could be exploited by malicious apps without any permission. This demonstrated the effectiveness
of our system and the limitation of the ad-hoc defense deployed by Google.

We further proposed a global reference counting mechanism to control
the number of Java Global Reference to defend the JGRE attack. The key observation
of this defense is that the number of Java Global Reference for a benign
app in the system service is typically small and in a reasonable range.
Thus we can safely set up a threshold of that number when creating a new
Java Global Reference. Then if the number exceeds the threshold, we deny
the creation of a new Java Global Reference so that the malicious app cannot
perform the attack to exhaust the resource. We evaluated our solution
to different Android versions, and it can successfully protect vulnerable services.

This paper makes the following contributions.

\begin{itemize}
    \item \textit{New discoveries.} We found that the latest Android version
    (Android 10) is still vulnerable, due to the incompleteness of the 
    ad-hoc defense mechanism proposed by Google.
    \item \textit{New tool and vulnerabilities.} We proposed a
    new tool called JGREAnalyzer, and applied it to various Android versions.
    It reported new vulnerabilities from Android 8.1 to Android 10.

    \item \textit{New Attacks.} We found multiple ways to attack the 
    Android system, even it keeps upgrading the JGRE defense. We studied the attack in-depth in the paper.
   \item  \textit{New Defense.} We proposed the defense mechanism to
    mitigate the JGRE attack. Our system defeats attacks that cannot be blocked by existing
    defense mechanisms.
    
\end{itemize}

\section{Background And Motivation}
In this section, we explain how Java Native Interface (JNI) is used in the Android framework, illustrate the vulnerable code by real example, and present our attacking methods with proof of concept (PoC) code.
\subsection{JNI Global Reference}

    The runtime of the Android system is built using both Java and native language
    (C and C++). To interact between the Java program and the native program,
    the Java Native Interface (JNI) \cite{jni_spec} is widely used, which allows the Java
    code to call or be called by the native code. Developers can implement the performance-sensitive parts in native code and provide flexible APIs in Java.
    When passing data between
    the Java program and native program, there is a special data structure
    called references that denote to Java objects inside the native program.
    There are two different types of reference, i.e., JNI Local Reference and JNI Global
    Reference. The difference between them is that the Java object denoted by the local reference could be garbage collected, while the Java object denoted
    by the global reference cannot be. JNI Global Reference(JGR) which is created via API NewGlobalRef is used for the native class sharing the same java object between different functions or threads. 
    
    We observe that Java global references (JGR) are widely used in many important modules such as Binder, Audio, OpenGL, Media, and USB drivers.

\begin{figure*}[t]
\center{\includegraphics[width=0.99\textwidth]{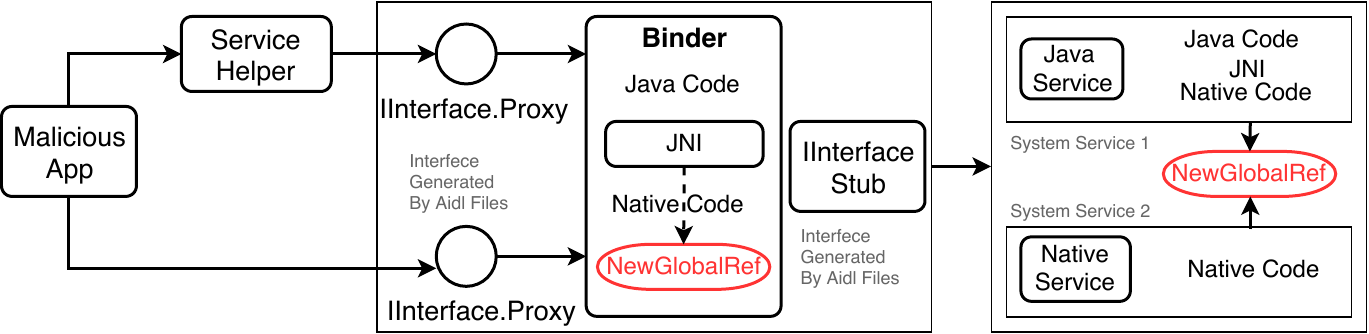}}
	\caption{The JGRE attack pattern}
	\label{jgre_attack}
\end{figure*}

    In the Android system, privileged operations are provided through Android
    services. These services expose different IPC interfaces
     that could be invoked by an Android app via the Binder~\cite{api_binder}. The IPC interfaces are defined in AIDL (Android Interface Definition Language) \cite{aidl} files which can generate Java IPC code for the service interfaces to communicate with the Binder. IInterface proxy classes are generated for the client process to send a request to the Binder and the Interface stub classes are 
     generated for calling the remote method implementations on the server-side. Other apps can invoke the service interfaces by using the IInterface proxy class or via service helper classes~\cite{gu16} which perform security checking and provide easy-to-use interfaces. However, the Binder IPC mechanism can be attacked by malicious apps via the JGRE attack. Android services are mainly implemented using native code (C++) and provide Java APIs for the framework layer via wrapper classes. JNI and JGR are widely used in native wrapper classes as these wrappers need to 
    access Java objects. It means that some service APIs need to create JGRs. As previously discussed, a JGR is not garbage collected by the runtime, which
    makes the system suffer from the DoS attack. Specifically, if an attacker can
    find a way to create many JGRs in the Java virtual machine, then these
    newly created objects will consume lots of memory, and overflow
    the JGR table - a special table that stores the pointers of JGRs. As a result, the process with overflowed
    JGR table will crash.  As is shown in Figure~\ref{jgre_attack}, the malicious app can exploit the vulnerable Service interface via IPC and create exceeding JGR via indirectly calling to NewGlobalRef and then make the JGR table overflow and services crashed. This type of DoS attack that consumes the JGR
    is called the JNI Global Reference Exhaustion (JGRE) attack~\cite{Gu2017JGRE}.
If the attacker is targeting core system services (e.g.,
\textit{system\_server}) in Android, it will lead to a soft reboot
as a fatal service has crashed, and the system cannot work before rebooting up that service again. 

In the JGRE attack, the malicious app calls Service API via Interface.Stub(). The Stub class implements the remote interface so that the user can use remote service in local, and deals with the data transferring between remote service and local. Importantly, it creates Java Global Reference. We present an example in the \textit{section 2.2} 

Since the exploitation of the vulnerability is through the exposed IPC interfaces of vulnerable system services, 
one potential way to mitigate the attack is by applying the permission-based access control mechanism of Android to IPC interfaces.
However, this does not work in practice, since it is tedious to define many permissions to confine 
every IPC interface (remember that legitimate
app that uses these interfaces have to request such permissions.) 
Google takes another way to add checks in many system service helper classes, such as notification service\cite{fixcode_notification} and WiFi service\cite{fixcode_wifi}. Service helper class usually can be considered as a reusable component and provide work that has no concrete business meaning whereas service class contains the business logic.

However,
some system services, such as the WiFi service, are still prone to the attack, even with the defense in WifiManager class\cite{fixcode_wifi}. 
Since Android version 9, Google proposed a generic defense mechanism for the JGRE attack, which sets a threshold of the number of binder proxy, and kills the process that exceeds the threshold. 
Original JGRE attack generates excessive JGR by generates excessive binder proxy, which usually creates at least one JGR. However, we still find ways to bypass the defense in both Android 9 and Android 10.

\subsection {A Real World Example}
\label{sec:vul_example}
Here we are presenting a motivation example of how JGR is used in the Android framework to explain why it may suffer from the JGRE attack. 
When a user wants to call the method \textit{startWatchingRoutes} in the \textit{audio} service, firstly, the user needs to create a Stub class by calling \textit{IInterface.Stub.asInterface()}, as shown in Figure \ref{fig:one_binder}. Then the user needs to call the method \textit{startWatchingRoutes} through this Stub, which will create a new IBinder object as well as calling the \textit{BpBinder:linkToDeath} method. From Figure \ref{newGlobalRef}, we can see how \textit{NewGlobalRef()} function, which generates JGR, is called. In the end, we construct multiple kinds of JGRE attacks by this calling method we just described.

\subsection {JGRE Attacks}
In the following section, we first review existing JGRE attacks, and then
propose a new JGRE attack, i.e., One Binder Attack that works towards the
latest Android versions (Android 9 and Android 10).

\subsubsection{Simple JGRE Attack}
\label{sec:attack_example}
\setcounter{footnote}{0}
 \textit{Simple JGRE attack} creates lots of JGR entries in the \texttt{system\_server} process by
constantly calling vulnerable interfaces. In particular,
we can use a vulnerable system service, i.e., audio service, to
launch the attack~\footnote{Note that, this service is running inside the system\_server process.}. 
We constantly invoke the vulnerable IPC interface \texttt{startWatchingRoutes()} of the audio service. 

\begin{figure}[t]
	\vspace*{-0.3cm}
	\setlength{\abovecaptionskip}{0.cm}
	\setlength{\belowcaptionskip}{-0.cm}
	\center
	\includegraphics{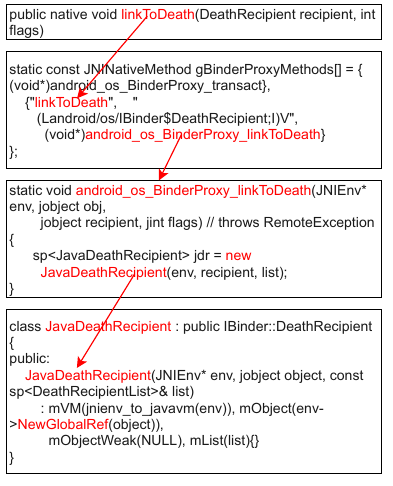}
	\caption{Calling path from linkToDeath(Java) to newGlobalRef(Native)}
	\label{newGlobalRef}
\end{figure}

Every time the interface is invoked, a new JGR
entry will be created. The background reason for JGR creating is that invoking vulnerable IPC interface will create a BpBinder object followed by calling BpBinder:linkToDeath, and linkToDeath calls NewGlobalRef following the path as shown in Figure \ref{newGlobalRef}. 
These JGR entries will not be released until the app
that invokes the interface exists. By doing so, we can make the \texttt{system\_server} process crash, and the Android framework will reboot.

\subsubsection{Service Based JGRE Attack}

Because malicious apps invoke vulnerable interfaces
to launch the Simple JGRE attack, Google added a defense mechanism to limit the total
number of invocations from one interface. Specifically, if the number
of Binder proxy, which is created through the invocation of the service
interface, exceeds a certain threshold (6,000), the app will be killed.
This simple defense can make the previous attack ineffective.
\begin{figure}[t]
	\vspace*{-0.3cm}
	\setlength{\abovecaptionskip}{0.cm}
	\setlength{\belowcaptionskip}{-0.cm}
	\center
	\includegraphics{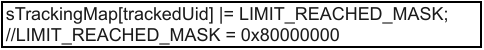}
	\caption{Code snippet for the Binder number limit. }
	\label{fig:counter}
	\vspace*{-0.3cm}
\end{figure}

However, we find a bug in the implementation of Android 9
(version before May 2018), which can be abused to bypass the defense.
Figure~\ref{fig:counter} shows the buggy code ($native/libs/binder/BpBinder.cpp$). The code uses a map 
(sTrackingMap) to track the number of Binder proxy for an
app (denoted by its trackedUid). When the system kills the app that has
exceeded the threshold, it will call the code shown in the snippet.
This code snippet sets the value to a substantial negative number, which
is around -2147467005.  

The whole attack consists of two stages. The first-stage attack creates a background service and then invoke the vulnerable interface as in the previous attack.
When the Binder proxy reaches the threshold, the app's service will be killed.
However, due to the bug in Figure~\ref{fig:counter}, the map used
to track the total number of the app is a big negative value. Then in
the second-stage attack, the malicious app restarts and creates a large number of JGR entries through multiple services. Since the initialized value of the number is a negative value, a malicious app can overflow the JGR table before this counter resets and being killed.
This attack leverages two background services to launch the attack, so
we call it a Service Based JGRE attack in this paper.

\subsubsection{One Binder JGRE Attack}

In the latest versions of Android 9 and Android 10, Google has fixed the bug
that can be exploited to launch the service based JGRE attack. However, we can still construct a new attack that can bypass the threshold limit of the Binder proxy.

\begin{figure}[t]
	\vspace*{-0.3cm}
	\setlength{\abovecaptionskip}{0.cm}
	\setlength{\belowcaptionskip}{-0.cm}
	\center
	\includegraphics{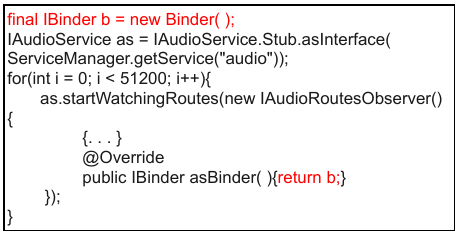}
	\caption{Code snippet for One-Binder attack. }
	\label{fig:one_binder}
	\vspace*{-0.3cm}
\end{figure}

Specifically, the malicious app can start a service and then invoke the
vulnerable interface as usual. However, during the attack, the app can change
the default value of the asBinder() function to return a final Binder object
instead of null. By doing so, we can still create JGR entries without
going through the checks added by Google. The background reason is that
the checks are inside the constructor function of BpBinder, and we can create
JGR entries without going through the constructor function. The code snippet in Figure ~\ref{fig:one_binder} shows part of the attack code.

The JGRE attack can be launched stealthily, causing the system into an unstable state without drawing attention from the user. For instance, the malicious app can register a listener to be automatically activated when the system boots. After that, it launches the attack to make the system reboot, which in turn actives the malicious app again. The system constantly reboots and the user hardly knows the reason. 
To our surprise, the JGRE attack can be launched remotely by using NativeScript \cite{NativeScript} and Cordova \cite{Cordova}. We found that the vulnerable system services can be called via JavaScript by using these frameworks, which means Android can be compromised by web apps. Android security is under dangerous situations as it can be attacked remotely.

\section{Design}
We propose JGREAnalyzer, a toolchain to detect and verify JGRE vulnerabilities in different versions of AOSP. In this section, we give an overview of JGREAnalyzer and present all the components including the static analysis module and dynamical analysis module in detail.

\subsection{Overview}
The root cause of the JGRE attack is that a malicious app could create lots of JGR through exposed IPC APIs. To detect the vulnerability, we propose a dynamic analysis module to automatically generate test code for each exposed IPC API. Then we monitor the execution of the code to determine whether the API is vulnerable by looking whether the system is rebooting or not. Unfortunately, there are two challenges to construct test code automatically: First, we need to create a binder object 
to connect to the service and it is difficult to get the service name from the service manager correctly.  
Second, the complicate parameters of these interfaces, e.g. \texttt{AlarmManager.cancel(OnAlarmListener)} takes an anonymous callback 
class \texttt{OnAlarmListener} as a parameter and sometimes 
the callback class needs other instances of the class to initial. The problem is some of these anonymous callback classes
need special parameters to construct e.g. some parameters can only be accessed via the singleton instances and some callback classes can not be accessed by third-party apps. As a result, dynamic analysis can not cover all the IPC methods. To solve the problem, we propose a static analysis module to analyze all the IPC methods and verify the results by dynamic analysis or test manually so that our tool can cover all the IPC methods.

\begin{figure*}[t]        
\center{\includegraphics[width=0.99\textwidth]{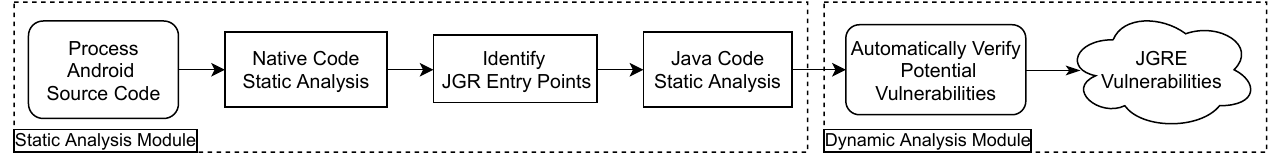}}
 \caption{The architecture of JGREAnalyzer}      
 \label{fig:design} 
\end{figure*}

The workflow of JGREAnalyzer is illustrated in Figure~\ref{fig:design}, and it contains a static analysis module and a dynamic analysis module. First of all, we process the AOSP source code and obtain the necessary files for further phases. Then we start the static analysis phase. In this step, we analyze both native code and Java code to identify potential JGRE vulnerabilities. Finally, we use a dynamic analysis module to validate potential vulnerabilities. We dynamically analyze all the IPC APIs more than the results of static analysis. 

Comparing to our previous work~\cite{Gu2017JGRE}, we design a new architecture and all the modules are rewrote, and hence the effects are significantly improved and most of the manual work can be done automatically. For static analysis, we implemented native code call graph analysis so that we do not need to find out the JGR entry methods manually. Moreover, we rewrite the Java code analysis module to automatically identify the JGRE vulnerable patterns as shown in section~\ref{sec:vul_example}. The dynamic analysis tool is also significantly improved. We propose a parameter engine that helps to generate correct parameters for service API so that the auto-generated test cases can run successfully. All these processes can run in the pipeline and the dynamic analysis tool can finish the verification of all the services API one Android version in 45 hours. Moreover, we evaluate the accuracy of our static analysis tool based on the results of dynamic analysis and manual verification.

\subsection{Static Analysis Module}
Since the JGRs are created in native code and invoked by Java JNI methods. The life cycle of JGR is managed by Java code. We use static analysis to find out the vulnerable IPC interfaces which could create JGR and keep references of the JGR. When these interfaces are exploited and continuously creating JGR in Binder or system core services, the runtime of Binder or core services crashes causing the system reboot due to JGR table overflow.

We perform call graph analysis for both native code and Java code. For native analysis, we need to find out the JNI methods which create JGR and then analyze the usages of these JNI methods in Java code. If it exists an execution path from service API in Java code to \texttt{NewGlobalRef()} to create JGR in native code and the JGR object is saved by the Java code, such as put the JGR object into a list, the service API may be vulnerable as the JGR object cannot auto release in native code via ref-counting when the Java object is not released by the Java GC. 

As illustrated in Figure~\ref{fig:design}, the static analysis operates in four steps, first, we compile the android source to get compile database for native analysis and extract an extend Jar for Java analysis. Then we use native call graph analysis to find out the JGR entry methods which are JNI methods that create JGR. We manually analyze these JGR entry methods and check the usages of these JNI methods to filter out these methods that are not running in the Binder process. Finally, we perform Java call graph analysis to find out the IPC interfaces which call these JNI methods in Binder and save references of these JGR object. We elaborate on it in the following sections.

\subsubsection{Processing the Android Source Code}
JGREAnalyzer works on various AOSP versions and uses both source code and compiling results as input. 
Source code is required as native code analysis needs to parse the raw CPP code and dynamic analysis also needs
to obtain the real name of parameters and local files in Java code which lost after compiling.
Our analysis targets are service API whose implementation code is distributed throughout the 
code-base (in various sub-directories of \texttt{aosp/frameworks}). Then the Soot based Java code analyzer chooses to use 
Jar as an analysis target to include all the whole services implementation and exclude useless third-party Java packages such as 
\texttt{org.apache.http} and we only need to analyze the code in the package \texttt{android.*} and \texttt{com.android*}. 
In addition, these Jar files can be obtained from real Android devices or emulators and make it possible for our Java analyzer to work on other customized Android versions such as Samsung or Huawei. In order to keep different
modules of our tool work on the same code-base, we choose to use the Jars from the compiling results which are consistent with the Source Code used by dynamic analysis. Consequently, our full toolchain can only work on open-source Android versions such as AOSP and Lingaros.

We compile the source code for two purposes: First, for Java static analyzer, we need to extract a combined Jar of \texttt{framework.jar} and \texttt{android.jar} which contain all the SDK code and framework code including exploitable hidden APIs.
Second, the clang based native code analyzer needs a compiled database for static analysis.

\subsubsection{Static Analysis for Native Code}
The goal of native static analysis is to find out the JNI methods which create JGRs and can be exploited by service APIs. We build a call graph for each JNI method and verify if the \texttt{NewGlobalRef()} method is called.

\noindent{\bf Call graph analysis for native code.} We get the native JNI methods for the Java JNI methods from Java static analysis via JNI relation mapping. These native JNI methods can be exploited via Java methods and are reachable methods for some Service APIs. Starting from these native JNI methods, we perform call graph analysis to find out whether these methods have invoked \texttt{NewGlobalRef()}. Our approach is based on libclang and requires the compiling commands exported from CMake so that clang can get the right type of the variables of the source code. Firstly, we parse and save the method definitions in all the C++ header files via traversing AST. After that, we resolve a method implementation in C++ source files only if the method is invoked by others. In this way, we do not need to process all the source files in the framework. We use the breadth-first search approach to build a call graph on-the-fly. Setting the reachable native JNI methods as entry points, we walk through the AST of each method and find out all the invoked methods. Then starting from the invoked methods, we locate the method implementation code by querying the definitions analysis results and parses the AST to find out invoked methods, and continues this procedure until we have gone through all the invoked methods or find the exiting point: \texttt{NewGlobalRef()} is invoked. Then we save the edges of the calling path which starts from an entry point to the exit point. The advantage of our approach is can minimize the call graph edge and process fewer C++ source files, which makes our tool more efficient.

The calling paths need to be further verified whether they are reachable as we do not perform control flow analysis to check the executing conditions. We manually confirmed these paths. Fortunately, the number is very limit. We start from all the native JNI methods and only find 26 paths in Android 9 and confirmed that all these paths can be reached on some conditions and we also make sure there are no missing exploited paths, such as Binder constructors and Binder:linkToDeath.

\begin{figure}[t]
	\vspace*{-0.3cm}
	\setlength{\abovecaptionskip}{0.cm}
	\setlength{\belowcaptionskip}{-0.cm}
	\center
	\includegraphics{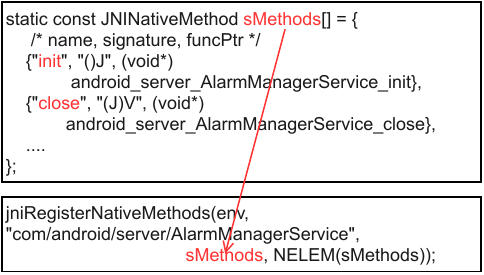}
	\caption{Code snippet for JNI register in alarm manager service. }
	\label{jni_register}
\end{figure}

\noindent{\bf Extracting JNI Relation Mapping.} 
JNI method binds a native method with a Java method name so that we can invoke the native method from Java code via Java JNI method. 
Since the JNI binding relations are defined and registered in native code,  
we need to analyze the native code to retrieve the JNI relation mapping of Java JNI methods to native JNI methods. 
Retrieving a complete JNI relation mapping is challenging as the definition of JNI methods is scattered in different files throughout the source code.
In Android, all the JNI modules are initialed in \texttt{AndroidRuntime.cpp} file when the Java Virtual Machine starts.
By analyzing the initialization of JNI modules, we can get a complete list of JNI files. 
However, we still need to parse these files to get the detailed definition of JNI methods as they are scattered in different files and registered via \texttt{jniRegisterNativeMethods} as is shown in Figure~\ref{jni_register}. 

The Java JNI method \texttt{init} from Java class \texttt{AlarmManagerService} is bind to native JNI method \texttt{android\_server\_AlarmManagerService\_init}. By resolving the Java class name, package name, and method name of each Java JNI method as well as its native JNI method from the JNI register code blocks, we gain the relation mapping of Java JNI methods to their native JNI methods.

Giving a Java JNI method, we can use the JNI relation mapping to get the corresponding JNI native method and perform call graph analysis in native code. After native code analysis, we obtain all the JNI methods which need to create JGR to save the reference of Java object. We regard these JNI methods as JGR entry methods. For each JGR entry method, we manually verify the correctness of its native call graph analysis and corresponding Java method name from JNI relation mapping as they are fatal for the further Java code analysis.

\subsubsection{Identify JGR Creating Methods}
Notice that only JGR created in core services via Binder may suffer DoS attack, we have obtained over 40 JGR entry methods and only 5 of them are used in Binder. We summary the JGR creating methods in Table~\ref{table:jgr_entry} and in the Java analysis we focused on these JGR creating methods. If a Java Binder object has invoked these JGR creating methods in service API, it may suffer JGRE attack.

\subsubsection{Static Analysis for Java Code}
Since vulnerable service APIs can be exploited via service AIDL interfaces or service helper interfaces. We need to analyze both service helper class in \texttt{android.jar} and system service classes in \texttt{framework.jar} which are the implementations of system services that can be invoked by AIDL interfaces. By combining all the code into an extended Jar file after compiling the source code, we can extract all the service helper classes and system service classes. Using the IPC interfaces of service helpers and service classes as entry points, we perform Java call graph analysis to find out the interfaces which call JGR entry methods. Then we check whether these JGR objects are saved in collections such as list, map, and set.

\begin{figure}[t]
	\vspace*{-0.3cm}
	\setlength{\abovecaptionskip}{0.cm}
	\setlength{\belowcaptionskip}{-0.cm}
	\center
	\includegraphics{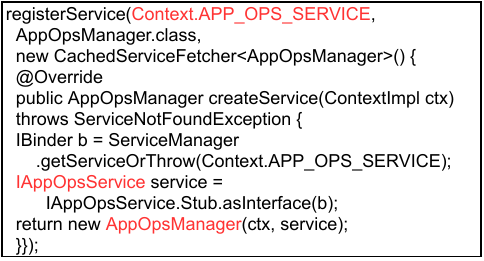}
	\caption{Code snippet for service helper classes register in SystemServiceRegistry class. }
	\label{fig:service_helper}
\end{figure}

\noindent{\bf Extracting target classes.} Service helper classes and system service classes are distributed in various packages, we analysis the register code blocks to resolve the class name and service register name of these services. All service helper classes need to be registered in \texttt{SystemServiceRegistry} class via \texttt{registerService(name, service)}.  As is shown in Figure~\ref{fig:service_helper}, we analysis the parameters of \texttt{registerService()} method and extract the return type of nested class's \texttt{createService()} method. 

In this example, we can obtain the return type \texttt{AppOsManager} which is a service helper class. However, sometimes the return value's type is not easy to obtain as it may be the result of function calls, we perform point-to analysis to get the real type of the return class. System services need to register to \texttt{ServiceManager} via \texttt{addService(name, service)} method. We locate all the \texttt{addService()} calls in framework code which are invoked in various classes. Moreover, the service name and implementation class can also be obtained by resolving the \texttt{addService()} method's parameters with point-to analysis. Note that, there are several services registered in native code and we ignore these services as our tools do not support native services.

Different from the Java program, the Android framework does not have a main method as entry points. We need to start with the public interfaces of services that can directly be called by third-party apps. Thus, all the public interfaces of services can be used as entry points. As is shown in Figure~\ref{jgre_attack}, two kinds of public interfaces need to be considered, we collect all the public interfaces of a service AIDL file which defines the app-accessible APIs of system services as well as public interfaces of service helper class which are all available to apps. After extracting the system service classes and service helper classes, we also extract the public interfaces of these classes and utilize them as entry points of the call graph analysis.

\noindent{\bf Call graph analysis for service APIs.} As is depicted in Section~\ref{sec:vul_example}, vulnerable APIs create JGR and save JGR Java object to collections which makes it unavailable to be released by Java Garbage Collector. We perform context-sensitive call graph analysis to identify the vulnerable patterns in service classes. Similar to native analysis, we also use a breadth-first search approach to build call graphs for each entry point method. To accelerate this process, as it is time-consuming to analyze all the call nodes of an entry method, the search depth is limited to at most 3 or 4 levels. Starting from the entry point method, for each method, we add all the methods invoked by it into the work list and then continue to analyze methods in work list. In this way, we analyze all the methods called by the entry point directly or indirectly, and the max analysis level is 4. The walk-through of call chains in most services methods can finish in 3 levels.

\begin{figure}[t]
	\vspace*{-0.3cm}
	\setlength{\abovecaptionskip}{0.cm}
	\setlength{\belowcaptionskip}{-0.cm}
	\center
	\includegraphics{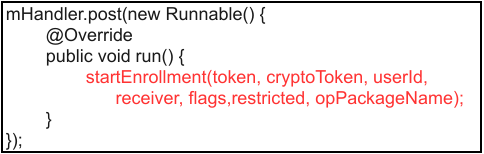}
	\caption{Code snippet for implicit control flow method in Handler class. }
	\label{fig:implicit_call}
\end{figure}

To improve accuracy, the virtual calls and implicit control flows need to be handled. We use CHA~\cite{cha_paper} (class hierarchy analysis) to resolve all the child classes of a virtual call method's declaring class and find the possible implementation of the method. There are two kinds of implicit control flow in system services, IPC calls, and some callback classes such as Handler and Messenger. For IPC calls, which are invoked via AIDL interfaces, we change the interface invokes to its implementation methods in the corresponding \texttt{IInterface.Stub} class. For other callback classes, we use the results of EdgeMiner~\cite{edgeminer} which marks all the implicit control flow methods. For instance, as is shown in Figure~\ref{fig:implicit_call}, when analyzing the call graph, we can only find \texttt{Handler.post()} method is invoked directly.

In fact, the \texttt{startEnrollment()} method will be invoked in another thread. By reading the method list from EdgeMiner, we can find \texttt{Handler.post()} is in it and know that it is a implicit control flow method. Therefore the methods of \texttt{Runnable} interfaces are invoked. Then we can get a \texttt{Runnable.run()} invoke statement here and do not miss the \texttt{startEnrollment()} call by analysis method calls in method \texttt{run()}.

The exit points of call graph analysis are collection add methods such as \texttt{list.add}, \texttt{map.put}. Note that, \texttt{RemoteCallbackList} class is also the most used collection for services. When we find a collection put operation, we check if the object put into the collection is an object which keeps a reference to JGR. When we find the collection is not a local object and save JGR related objects, the JGR related object can not be cleaned by the Garbage Collector and may suffer JGRE DoS attack, the call graph analysis will exit here, otherwise, continue to analyze all the method calls in call graph until the analyze level reaches 4. Moreover, to identify the real parameter type of collection put invokes whose type is \texttt{Object} in compiling code, we perform context-sensitive point-to analysis to obtain the real type (Class) of values which putting into collections by recording the definition of each value and also need to record the pointer reference of values in assign statements. Firstly, we precisely get the parameter object's type. Secondly, we determine if it is JGR related object by verifying if it is Binder class or if it contains fields which are Binder classes. 

After analyzing all the system service interfaces, we can use the results to analyze service helper interfaces that wrapped the system service interfaces and finally called them via IPC interfaces generated by AIDL files. By checking if these service helper interfaces have invoked vulnerable system service interfaces, we can quickly find out vulnerabilities in service helper interfaces.

\subsection{Dynamic Verification Module}
After static analysis, we discover the potential vulnerable Service APIs and their executing paths which could finally be exploited by third-party apps. As some vulnerabilities can not be exploited due to  private API or requiring system level permission. We use a dynamic analysis module to automatically generating PoC apps for these vulnerabilities and check if these apps can perform DoS attacks on real Android devices.

\subsubsection{PoC App Generation}
As is elaborated in Section~\ref{sec:attack_example}, the PoC code needs to get the Binder object of services using the service register name, e.g. the name of \texttt{AudioService} is \textit{audio}, which can be gained by the static analysis module. We also need to feed parameters to the service interfaces, which is a critical challenge to generate the correct parameters. These parameters are range from base value types such as \textit{int} and \textit{bool} to class types including interface callbacks and singleton instance. 

To address the problem, we use Java Parser~\cite{java_parser} to parse the source code of AOSP to generate a method definition model in the JSON format for each service interface, which includes the class name, the method definitions, the dependency packages and permissions, the parameters, and the return type of the method. In particular, we set some rules to generate  parameters with special names such as \textit{package} and \textit{uid} which means user identity names and we feed corresponding values from a set of packages name and user Id. For singleton instances, such as \textit{looper}, \textit{context}, we also use preset values. The core idea is collecting raw information from source code to generate classes override the callback interfaces and manually set some default values for the special parameters based on name and type. 
The PoC generation module changes different values for parameters and tries to compile the app. All the service interfaces process in pipelines and all the compiling results will be recorded. We manually check the failed ones and improve the special parameter generation phase.  However, there is still some PoC code that failed to compile as the interface or parameter is private or in interface blacklist.

\subsubsection{Vulnerability Verification}
\label{parameter_engine}
The dynamic verification tool runs the compiled PoC app in real devices via adb command and keeps monitoring the log of logcat. When it discovered the core dump message as the reason for JGRE, it means the PoC app success to perform DoS. We mark the interfaces of these PoC apps as vulnerable APIs and record which the Android versions it can attack successfully.

\section{Implementation}
\label{sec:implementation}
JGREAnalyzer is implemented based on Soot~\cite{soot} and LibClang~\cite{libclang}. Its source code can be accessed on github\footnote{https://github.com/fripSide/JGREAnalysis}.

\noindent{\bf Processing android source code.}
We compile the AOSP source code to extract extended Jar files which contain all the
service implementation and gain the compiling database of Clang from the native code. 
It is not trivial to extract the Jar files as different Android versions use various
compiling strategies and unique dex formats. For versions before Android 8, Android
adopts \textit{odex} as dex file format and we need to dump all bytecode including
 \textit{odex}, \textit{oat} and 
other bytecode files from a specific Android image built by ourselves. Then we use the
oat2dex~\cite{oat2dex} tool to convert these files into dex files. At last,
the dex files are processed into jar files by the dex2jar~\cite{Dex2jar} tool and we
repackaged these jar files to get an extension \texttt{framework.jar}. For the later
versions after Android 8.0, android no longer utilizes the \textit{odex} files, 
instead a new format named \textit{vdex}~\cite{how_art_works} is used. 
VdexExtractor~\cite{VdexExtractor} can be used to translate the \textit{vdex} files to
\textit{dex} files which can be processed by dex2jar. We extracted the extension framework jar 
files from Android 6 to the Android 9 but failed to process the newest Android 10. 
The jar files and our detailed approaches can be found in Github~\cite{AndroidApiExtract}. 

\noindent{\bf Static analysis for native code.} 
Libclang requires a compiling database to retrieve the file including references. Thus, we compile
the source code and generate the \texttt{compile\_commands.json} via the CMake commands. 
Then we parse the AST of header files and identify the method declaration by clang
cursor type including \texttt{CLASS\_DECL} nodes which contain class methods and \texttt{FUNCTION\_DECL} 
nodes which define a normal function to
collection of the method definitions list.
After gathering all the methods definitions, we 
start from native JNI methods which are entry points and parse the invoke expressions including
\texttt{CALL\_EXPR} which is method call nodes and the class initializes nodes which invoke the constructor of classes. 
For the invoke expressions, we can locate their method implementation from the method definitions list and then parse the
definition files until we have finished all the invoked expressions or stop at the \texttt{NewGlobalRef()} method. 
We save the calling paths and manually confirm them by reading the source code.

\noindent{\bf Static analysis for Java code.} We implement Java static analysis based on Soot~\cite{soot_url} and use Jimple code as an intermediate representation(IR). A challenge is to retrieve the real class type of a parameter of a method call as the parameter type may be marked as a parent type such as \texttt{IBinder} or \texttt{Object} in the method signature. In Java, the real type of a class is dynamically dispatched and we perform point-to analysis to resolve the real type of a method parameter. We track the assigned statements of the method arguments in the various procedure and finally locate the definition of the value. This technology is used to identify the real class type of system services and the collection put operation.  For call graph building, we do not use soot's builtin CHA or spark call-graph as we do not need to build a full call graph by performing whole program analysis based on the sight that only a small part of the framework is related to service APIs. We build a call graph by walking through the methods invokes by each service interface and exit and return immediately when we find the JGR entry methods so that the rest of the invokes will not be analyzed. We also limit the max search levels as in practice, almost all the vulnerable APIs are invoked directly in the service interface implementation method and this approach can save about 30\% analysis time. 

\noindent{\bf Dynamic verification of vulnerabilities.}
The automatic verification tool writes the PoC code by leveraging JavaPoet~\cite{javapoet} and copy the PoC code to the template app project. Then it tries to use different parameters for compiling the PoC code for each interface. If the code is compiled successfully, the apps will be installed and run on Android devices via ADB commands. The executing log can be accessed via the ADB tool and we save the log for each app. All the PoC apps are running in the pipeline and sometimes the system blocks and app is installed failed. We need to reboot the devices automatically or manually and then continue to verify the rest apps. At last, all the JGRE error logs are extracted and the vulnerable interfaces are reported. 
\section{Evaluation}
In this section, we evaluate the accuracy and effectiveness of JGREAnalyzer. We study five Android versions range from Android 6 to the newest Android 10 and discuss the detail of vulnerabilities and Google's countermeasures in different versions. All the experiments are performed on a laptop computer with a 6-core Intel CPU (i7-8750H) and 16GB memory with Ubuntu $18.04$. 

\begin{table}[t]
	\centering
	\caption{The statistics of market share and service API number in different Android versions. In the third column, the two numbers are the counts of public interfaces in service helper class and in system services.}
	\begin{tabular}{llll}
			\toprule
			Version & Market Share   & Service APIs & Service Classes \\
			\midrule
			Android 6.0     & 16.9\%  & 1776/2031  &  68/79  \\ 
			Android 7.1     & 7.8\%   & 1755/3709 & 74/102  \\ 
			Android 8.1     & 12.9\%  & 2537/3118 & 85/113   \\ 
			Android 9.0     & 10.4\%  & 2284/3363 & 93/121   \\ 
			Android 10.0    & \textless{}0.1\% & -  & -  \\  
			\bottomrule
	\end{tabular}
	\label{table:target}
\end{table}

\noindent\textbf{Analysis target.} We apply the static analysis tool on each Android version and shown in Table~\ref{table:target}. For different Android versions, the service number increase as the version number and the vulnerability number decrease as Google continues to fix these problems. It shows that the old Android versions still has a very large market share~\cite{market_share}. Even if the Android team has fixed the attacks in JGRE~\cite{Gu2017JGRE} since Android 7, but our new attack method still can work on these versions and all the Android versions are affected.  Note that, we failed to extract the \texttt{framework.jar} for Android 10 as it have adopted a new VM bytecode format which is different to previous Android versions. Consequently, our static analysis tool cannot process Android 10 but we still verified the existing vulnerabilities on Android 10 with our new attack methods.

\noindent\textbf{Tool Efficiency.} The static analysis module of JGREAnalyzer can finish analyzing each version of Android extend Jar in less than 15 minutes and finish native code analysis in less than 5 minutes. The dynamic verification is time-consuming which needs about 40 hours for each Android version, but it is acceptable comparing to the time-frame of Android major version updates, which is in years.

\subsection{Tool Accuracy}
After analyzing 5 Android versions, JGREAnalyzer static analysis module total report 277 vulnerabilities. We verify the analysis results of Android 9 to evaluate the accuracy of our tool. The results of the static analysis module including native analysis and Java analysis are manually verified respectively. The accuracy of the automatic dynamic verification tool can be regarded as 100\% as the JGRE DoS attacks can be precisely captured by analyzing the system crash log. Hence, we only discuss the effectiveness of it.

The native code analysis module of JGREAnalyzer discover $24$ native JGR creating methods and we manually verified all the JNI methods in Android 10. We find 29 JGR creating methods in total including the $24$ methods discovered by the static analysis tool. We analyze the false negative case (missing reports) and find out the reasons. Finally, we construct Java code to verify if the calling path of all these native methods can be triggered as the executing conditions are complicated to verify by only reviewing the code. As shown in Table \ref{table:jgr_entry}, we find most of these JGR creating methods are hidden methods that can not be exploited by third-party apps and only 4 methods can be used to perform DoS to app or system. 

\begin{table}[t]
	\centering
	\caption{JGR creating methods in native code.}
		\begin{tabular}{ll}
		\toprule
			Module & JGR Creating API in Binder  \\ 
		\midrule
			Binder     &  android\_os\_BinderProxy\_linkToDeath    \\ 
			           &  android\_os\_BinderProxy\_unlinkToDeath    \\
			Other      &  android\_media\_AudioRecord\_setup    \\ 
			           &  android\_media\_JetPlayer\_setup \\ 
			           &  android\_media\_AudioTrack\_setup \\
			           &  android\_hardware\_UsbRequest\_queue\_array    \\ 
			           &  android\_hardware\_UsbRequest\_queue \\ 
			\bottomrule
	\end{tabular}
	\label{table:jgr_entry}
\end{table}

Only 2 methods in Binder can make the core service crash and cause system reboot. APIs like \texttt{SurfaceTexture::init()} only make the app crash due to JGRE as it runs in the third-part app process. 

After Java static analysis, we find out 73 potential vulnerabilities. However, not all these vulnerabilities as some need system-level permission or use hidden APIs which may not be exploited by third-party apps.  The accuracy of the Java analysis module report 45 vulnerable methods in total. After dynamical verification of the vulnerabilities, we find out 35 APIs can be exploited in real devices. We do not regard the non-exploitable APIs as the false positives as these results may be true in a static analysis view. Some of these hidden apps can be exploited by bypassing the system checking.  Unfortunately, the false-negative rate is impossible to statistic as there may be 0-day bugs of DoS to system service~\cite{cve_new} and we can not ensure the vulnerabilities that we discovered are complete. Other service vulnerability static analysis tools (e.g. Invettor~\cite{invetter}, Kratos~\cite{ss_attack_2}) also do not evaluate the completeness and the false negatives for the same reason. We manually verify the code of the interfaces which failed to exploit and find that 4 interfaces are false-positive cases due to special control flow, e.g. creating and adding JGR only when the object does not exist. We also find 2 false-negative cases that proved to can attack but do not report the static analysis tool. The overall true positive rate of static analysis is 94.5\%(69/73). The 69 vulnerabilities report by the static analysis can create unlimited numbers of JGR and keep the reference of these JGR.

\subsection{Tool Effectiveness}
The results of static analysis in different Android versions are shown in Table~\ref{table:vul_version}. 

\begin{table}[t]
	\centering
	\caption{Vulnerabilities in different Android versions.}
	\begin{tabular}{llll}
            \toprule
			Android Version & All & TP & Exploitable \\ 
			\midrule
			Android 6(api 23)      &  80  & 75 & 54 \\ 
			Android 7.1(api 25)      &  59  & 54 & - \\ 
    		Android 8.1(api 27)     &  71  & 65 & 35 \\ 
    		Android 9(api 28)      &  73  & 68 & 23  \\ 
    		Android 10(api 29)      &  -  & - & 21  \\ 
    		\bottomrule
	\end{tabular}
	\label{table:vul_version}
\end{table}

In total, JGREAnalyzer report 75 different vulnerabilities and we further verify if these vulnerabilities are exploitable. After dynamic verification, we find that 35 of these vulnerabilities can be exploited on at least one major Android version. The left APIs are vulnerable in code but can not be exploited as they are private or hidden to third-party apps. Admittedly, these APIs may be vulnerable as there are other methods to exploit them such as bypassing the greylist~\cite{gray_list} and access the hidden API~\cite{hidden_api}. The root causes of all these vulnerabilities are the same: the native API \texttt{linkToDeath()} in Binder is not protected and can be exploited to creating unlimited JGR without releasing. Thus, we do not make more efforts to confirm all these potential vulnerable Java APIs. 

To evaluate the effectiveness of the dynamic verification module, we use JGREAnalyzer to generate test apps for all the public service APIs including all the vulnerable APIs report by the static analysis tools. It can automatically verify 64\% of these methods (720/1120) and find out 35 vulnerabilities. There are two extra vulnerabilities that are not reported by the static analysis. We also find out some APIs can make the system blocking with no response to ADB and need to reboot, such as \texttt{playSoundEffect} which can create too many floating windows. To further assess how these vulnerabilities affect different Android versions, we use all the attack methods in \ref{sec:attack_example} and test these vulnerabilities in different Android versions shown in ~\ref{table:results}. Since Android 7.1 has no big JGRE security updates from Android 6, we did not test how many interfaces were exploitable on Android 7.1. 

JGREAnalyzer has a reasonably high true positive rate, ranging from 91.52\% to 93.75\%, in finding JGR vulnerabilities in Android. Surprisingly, after 4 major versions of updates, we still found 16 vulnerabilities in the newest Android version.
We submit all these vulnerabilities to Google as they only receive reports on the latest Android. The problem is confirmed by Google with Android Bug ID 141758688.

\subsection{Analysis Results}
\label{sec:results}
Comparing to JGRE~\cite{Gu2017JGRE}, we find 12 new extra vulnerabilities. 23 of the vulnerabilities report by JGRE are still vulnerable with new attack methods and the APIs other than these 23 vulnerabilities in JGRE can not work in the later Android version. Using Android 8.1 as the baseline, we find 35 exploitable vulnerabilities in system service APIs shown in Table~\ref{table:results}.  Finally, we analyze the service helper APIs in Android 10. 

\begin{table*}[t]
    \centering
    \caption{Vulnerable IPC Interface Analysis Results}
	\begin{tabular}{|c|l|l|l|}
		\hline
		\multirow{2}{*}{Service Name}       & \multirow{2}{*}{Vulnerable IPC Interface} & \multicolumn{1}{c|}{Afftected AOSP} & \multicolumn{1}{c|}{\multirow{2}{*}{In GreyList}} \\ \cline{3-3}
		&                                           & \multicolumn{1}{c|}{\ 6 \quad 8.1 \quad 9 \quad 10 \ }  &    \\ \hline
		\multirow{2}{*}{accessibility}      & addAccessibilityInterationConnection      & $\ \bullet \ \quad \bullet \ \; \quad \circ \ \; \quad \circ$ & $\qquad no$                        \\ \cline{2-4} 
		& addClient                                 & $\ \bullet \ \quad \bullet \ \; \quad \bullet \ \; \quad \bullet$  & $\qquad no$   \\ \hline
		activity                            & registerReceiver                          &  $\ \bullet \ \quad \bullet \ \; \quad \circ \ \; \quad \circ$ & $\qquad yes$                     \\ \hline
		\multirow{2}{*}{appops}    & startWatchingActive                 & $\ \bullet \ \quad \bullet \ \; \quad \bullet \ \; \quad \bullet$ &            $\qquad no$            \\ \cline{2-4}                          
		& startWatchingMode                         & $\ \bullet \ \quad \bullet \ \; \quad \bullet \ \; \quad \circ$ & $\qquad yes$                        \\ \hline
		\multirow{3}{*}{audio}              & registerPlaybackCallback                  &$\ \bullet \ \quad \bullet \ \; \quad \bullet \ \; \quad \bullet$ & $\qquad no$                       \\ \cline{2-4} 
		& registerRecordingCallback                 & $\ \bullet \ \quad \bullet \ \; \quad \bullet \ \; \quad \bullet$ & $\qquad no$                      \\ \cline{2-4} 
		& startWatchingRoutes                       & $\ \bullet \ \quad \bullet \ \; \quad \bullet \ \; \quad \bullet$  & $\qquad yes$                      \\ \hline
		autofill                           & addClient                                 & $\ \bullet \ \quad \bullet \ \; \quad \bullet \ \; \quad \bullet$ &   $\qquad no$                      \\ \hline
		clipboard                           & addPrimaryClipChangedListener             & $\ \bullet \ \quad \bullet \ \; \quad \bullet \ \; \quad \bullet$ & $\qquad no$                       \\ \hline
		\multirow{2}{*}{content}    & addStatusChangeListener                 & $\ \bullet \ \quad \bullet \ \; \quad \bullet \ \; \quad \bullet$ &   $\qquad no$                    \\ \cline{2-4}                          
		& registerContentObserver                         & $\ \bullet \ \quad \bullet \ \; \quad \bullet \ \; \quad \bullet$ & $\qquad no$                \\ \hline
		country\_detector                   & addCountryListener                        & $\ \bullet \ \quad \bullet \ \; \quad \bullet \ \; \quad \bullet$ & $\qquad yes$                       \\ \hline
		deviceidle                          & registerMaintenanceActivityListener       & $\ \bullet \ \quad \bullet \ \; \quad \bullet \ \; \quad \bullet$ & $\qquad no$                                             \\ \hline
		ethernet                            & addListener                               & $\ \bullet \ \quad \bullet \ \; \quad \bullet \ \; \quad \bullet$ & $\qquad no$                                             \\ \hline
		fingerprint                         & addLockoutResetCallback                   & $\ \bullet \ \quad \bullet \ \; \quad \bullet \ \; \quad \bullet$ & $\qquad no$                                            \\ \hline
		input\_manager                      & vibrate                                   & $\ \bullet \ \quad \bullet \ \; \quad \bullet \ \; \quad \circ$ & $\qquad no$                                            \\ \hline
		location                      & addGpsStatusListener                                   & $\ \bullet \ \quad \circ \ \; \quad \circ \ \; \quad \circ$ & $\qquad no$                                            \\ \hline
		media\_router                       & registerClientAsUser                      & $\ \bullet \ \quad \bullet \ \; \quad \bullet \ \; \quad \circ$ & $\qquad no$                                             \\ \hline
		\multirow{2}{*}{media\_session}     & registerCallbackListener                  & $\ \bullet \ \quad \bullet \ \; \quad \circ \ \; \quad \circ$ &  $\qquad no$                                            \\ \cline{2-4} 
		& createSession                             & $\ \bullet \ \quad \bullet \ \; \quad \circ \ \; \quad \circ$ & $\qquad no$                                             \\ \hline
		\multirow{2}{*}{midi}               & openBluetoothDevice                       & $\ \bullet \ \quad \bullet \ \; \quad \circ \ \; \quad \circ$ & $\qquad no$                                             \\ \cline{2-4} 
		& registerDeviceServer                      & $\ \bullet \ \quad \bullet \ \; \quad \circ \ \; \quad \circ$ & $\qquad no$                                             \\ \hline
		mount                 & registerListener           & $\ \bullet \ \quad \circ \ \; \quad \circ \ \; \quad \circ$ &  $\qquad no$                                            \\ \hline
		network\_management                 & registerNetworkActivityListener           & $\ \bullet \ \quad \bullet \ \; \quad \bullet \ \; \quad \bullet$ &  $\qquad no$                                            \\ \hline
		\multirow{3}{*}{print}              & print                                     & $\ \bullet \ \quad \bullet \ \; \quad \bullet \ \; \quad \bullet$ &  $\qquad no$                                            \\ \cline{2-4} 
		& addPrintJobStateChangeListener            & $\ \bullet \ \quad \bullet \ \; \quad \bullet \ \; \quad \bullet$ & $\qquad yes$                                             \\ \cline{2-4} 
		& createPrinterDiscoverySession             & $\ \bullet \ \quad \bullet \ \; \quad \bullet \ \; \quad \bullet$ & $\qquad no$                                             \\ \hline
		storage\_manager                 & registerListener           & $\ \bullet \ \quad \bullet \ \; \quad \bullet \ \; \quad \bullet$ & $\qquad no$                                             \\ \hline
		\multirow{3}{*}{telephony.registry} & addOnSubscriptionsChangedListener         & $\ \bullet \ \quad \bullet \ \; \quad \bullet \ \; \quad \circ$& $\qquad no$                                             \\ \cline{2-4} 
		& listen                                    & $\ \bullet \ \quad \bullet \ \; \quad \bullet \ \; \quad \circ$ & $\qquad yes$                                             \\ \cline{2-4} 
		& listenForSubscriber                       & $\ \bullet \ \quad \bullet \ \; \quad \bullet \ \; \quad \circ$ & $\qquad no$                                             \\ \hline
		wallpaper                           & registerWallpaperColorsCallback           & $\ \bullet \ \quad \bullet \ \; \quad \bullet \ \; \quad \bullet$ & $\qquad no$                                             \\ \hline
		\multirow{2}{*}{wifi}               & acquireWiFiLock                           & $\ \bullet \ \quad \bullet \ \; \quad \bullet \ \; \quad \bullet$ & $\qquad no$                                             \\ \cline{2-4} 
		& acquireMulticastLock                      & $\ \bullet \ \quad \bullet \ \; \quad \bullet \ \; \quad \circ$ & $\qquad no$                                             \\ \hline
		window                 & watchRotation           & $\ \bullet \ \quad \circ \ \; \quad \circ \ \; \quad \circ $ & $\qquad yes$                                             \\ \hline
		window\_manager                 & registerWallpaperVisibilityListener           & $\ \bullet \ \quad \bullet \ \; \quad \bullet \ \; \quad \bullet$ & $\qquad no$                                             \\ \hline
	\end{tabular}
	\label{table:results}
\end{table*}

In the following section, we summarize all the vulnerabilities discussed in JGRE~\cite{Gu2017JGRE} and our current results shown in section~\ref{sec:results} and explore the fixed vulnerabilities and the vulnerabilities which are still valid or emerging in the latest Android versions.

\subsubsection{Vulnerabilities fixed by Google}
As shown in Table~\ref{table:results},  since SDK 28, Google starts removing some previously discovered vulnerable interfaces from usable interface list. Google's defense method is not effective, even it sacrifices usability. Android system still under the JGRE threats. 

\subsubsection{Existing Vulnerabilities with new Attack Method}
Although Google set a limit for Binder Proxy to control the number of JNI Global Reference, we can use Service Based JGRE Attack and one Binder JGRE Attack to bypass the limit. \texttt{addAccessibilityInterationConnection} method of \texttt{accessibility} service is an interface that can only attack Android 8.1 and older versions and cannot be called by a user because Android limited this interface since SDK 28 (Android 9).
It is a special interface that can be exploited without lowering the complied SDK version, which means it can compile in SDK version 29 and can compromise every Android version. 
\texttt{audio.startWachingRoutes} can be used to attack every Android Version as long as it complied SDK version is 26, which is the minimum requirement SDK version set for Google Play.

\subsubsection{New Vulnerabilities}
New vulnerabilities are bringing in since the Android version increases. For instance, we found \textit{accessibility.addClient} and \textit{appops.startWatchingMode} in Android 9, and \textit{storage\textunderscore manager.registerListener} in Android 10. By using JGREAnalyzer, we found 12 vulnerabilities more than our previous work\cite{Gu2017JGRE}, many of which can be exploited from Android 6 to Android 10. 

\section{Countermeasure}
In this section, we revisit the defenses enabled in the current Android versions and then propose new defenses to defeat the JGRE attack.  

\noindent \textbf{Defense on Android 6.0.1} Google named JGRE vulnerability as a "resource exhaustion issue" in Android 6.0.1, and selectively fixed a small number of the vulnerable system services, such as Notification service \cite{fixcode_notification} and WiFi services \cite{fixcode_wifi}. They have been utilized several ad hoc defenses trying to fix the JGRE vulnerabilities. For example, the ad hoc defenses are that the system checks the JGR number in system services and system service helper. However, this defense method has two major challenges. First, it is challenging to choose specific thresholds suitable for all apps and all devices due to the fragmentation of the Android system. The thresholds vary in different interfaces because different interfaces provide different functions. In Android 6.0.1, Google sets the threshold to 50 per-process in \textit{WifiManager}, and sets the threshold to 1 per-process in \textit{InputManagerService}. The threshold needs to be set precisely to ensure that the Android system has high usability. Second, an attacker can utilize one vulnerable interface in any system service to attack the system process, since most system services run in \textit{system\textunderscore server} and share one JGR table.

\noindent \textbf{Defense on Android 8.1} Google still performs ad hoc defenses on Android 8.1 in two ways. One of the ways is that they modified some vulnerable interface methods so that an attacker cannot access these methods. The other way is elevating the permission level that the attacker needs to acquire. However, modifying interface methods, and elevating the permission level would hinder Android system usability to some extent. Moreover, there are still many vulnerable interface methods that can be utilized by an attacker. As Table \ref{table:results} shows, some interfaces were fixed by Android 8.1 defenses(We did not include all of them because of the page limitation.) However, there are 34 vulnerable IPC interfaces can be used to attack Android 8.1 directly.

\noindent \textbf{Defense on Android 9} Android 9 has a generic defense mechanism other than ad hoc defenses. Android 9 strictly controls the number of binder proxy. In the early version of Android 9, if the number of binder proxy of one app is more than the threshold, which is 2500 \cite{fixcode_2500bp}, the system will kill the app. However, sometimes SysUi will restart because the threshold is related low. Hence Android team decides to increase the limit to 6000 to reduce the chance of a SysUi restart \cite{fixcode_6000bp}.
Nevertheless, Android 9 still may kill benign apps, which create more than 6000 binder proxies in some circumstances. Unfortunately, with such strict binder creation policy, which even sacrifices usability for security, the defense mechanism of Android 9 still can be bypassed by Service Based Attack. As we mention before, there are 28 vulnerable IPC interfaces that can be used by One Binder JGRE Attack to compromise Android 9 system.
We need to mention that even though Google constrains some interfaces by adding them to the \textit{hidden-api-greylist}, we can still lunch JGRE attacks using some of the APIs in the greylist.

\noindent \textbf{Defense on Android 10} Android 10 develops a generic defense mechanism by constraining the number of binder proxies. Particularly, it can defend the Service Based Attack that can be used to exploit Android 9. However, Android 10 is still vulnerable to one Binder attack, which creates JGR through one binder and bypasses the mechanism in Android 10. Thus, there  still exist 21 vulnerabilities in Android 10.

\begin{figure}[t]
	\vspace*{-0.3cm}
	\setlength{\abovecaptionskip}{0.cm}
	\setlength{\belowcaptionskip}{-0.cm}
	\center
	\includegraphics{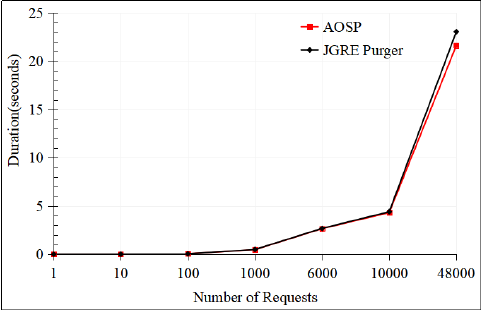}
	\caption{The Performance Overhead of Creating JGRs}
	\label{fig:overhead}
\end{figure}

\noindent \textbf{JGRE Purger} In order to complete the two major challenges in defeating JGRE attacks, we propose a new generic defense mechanism, JGRE Purger, which solves JGRE problem in native layer limiting the creation of a native object in JNI layer. 
Instead of detecting malicious apps, our system enforces the protection in the framework by adding a wrapper of the NewGlobalRef() function. As a result, we can keep track of the number of JGR of one particular app, as well as the total number of JGR. When a malicious app creates lots of JGR, our system will detect it since the number of JGR exceeds a threshold. In our system, this threshold is 6000 according to Gu et al.\cite{Gu2017JGRE}. As shown in Table \ref{table:overhead} and Figure \ref{fig:overhead}, the overhead of the JGRE Purger is very negligible.
Different from the defense mechanism from Google, JGRE Purger directly limits the creation of JGR, instead of controlling the in-directed creation of JGR (BpBinder) or limiting the usability of user-end IPC interfaces.

\begin{table*}[t]
 \centering
 \caption{The Performance Overhead of Creating JGRs}
\begin{tabular}{llllllll}
\toprule
 \diagbox[width=12em,trim=l]{Defense}{Duration}{Requests}           & 1  & 10 & 100 & 1000 & 6000 & 10000 & 48000 \\ 
 \midrule
JGRE Purger & 0.0016s & 0.0063s & 0.0476s  & 0.4986s   & 2.6812s   & 4.4077s    & 23.0409s   \\ 
AOSP        & 0.0018s & 0.0071s & 0.0502s  & 0.4827s   & 2.6585s   & 4.3148s    & 21.6134s   \\ 
\bottomrule
\end{tabular}
\label{table:overhead}
\end{table*}
\section{Discussion}

In this paper, we propose an approach that can both analyze Java code and native code to detect JGRE vulnerabilities. We first perform reachable analysis and then use points-to analysis to find out the usages of sensitive APIs. Comparing to other technologies such as symbol execution, our approach can precisely locate the vulnerabilities with a high efficiency. The analysis results show the root causes of the  vulnerabilities which help us understand this problem.

JGRE vulnerabilities exist on every Android version even though Google has fixed several times. The problem is that system service is modified and iterated in a very high frequency. We do not pay much attention to the prebuilt apps vulnerabilities \cite{Gu2017JGRE}, because we are proposing a generic defense mechanism, JGRE Purger. It has two advantages. The first advantage is that it enables generic protection against JGRE attacks in all IPC channels. Since it constrains the creation of JGR from the source, NewGlobalRef(), it can protect various IPC channels other than the Binder, such as IPC based on Linux sockets,
pipes, and signals. The second advantage is that it is agile with various versions of the Android system, because it provides a generic defense solution and the Android system has the characteristics of the fragmentation due to the
numerous combinations of different device models and Android versions.

\section{Related Work}
\textbf{Android analysis tools.} Static and dynamic analysis technology has been well researched, and researchers presented lots of sophisticated tools \cite{app_2}, \cite{feng2014apposcopy},  \cite{ma2016libradar}, \cite{ma2016libradar}, \cite{li2015iccta}. As the state-of-art static taint analysis tool, Flowdroid \cite{app_2} can be used to detect possible sensitive information leakage in Android apps. Apposcopy \cite{feng2014apposcopy} can recognize a common class of Android malware that steals user data by detecting if a particular application matches a malware signature. LibRadar \cite{ma2016libradar} is capable of finding third-party libraries in a particular Android app.  IccTA \cite{li2015iccta} can detect inter-app based privacy leaks by reproducing the context between the components of Android applications. 

\noindent \textbf{JNI security.} Many researchers studied JNI security \cite{Gu2017JGRE}, \cite{li2009finding}, \cite{chisnall2017cheri}, \cite{qian2014tracking}. Gu et al. \cite{Gu2017JGRE} analyze the JNI Global Reference resource exhaustion vulnerabilities in the Android system. In order to look at exceptions and report errors in JNI programs, Li and Tan \cite{li2009finding} proposed a unique static analysis framework.  Chisnall et al. \cite{chisnall2017cheri} show that they had implemented several security improvement methods for JNI. Qian \cite{qian2014tracking} check the interactions and trace information flows going through JNI in Android apps to detect information leakage.

\noindent \textbf{Service security and DoS attacks.} There are many DoS attacks that abuse vulnerabilities of Android system service \cite{ss_attack_1}, \cite{Gu2017JGRE}, \cite{cao2015towards}, \cite{invetter}, \cite{armando2012would}, \cite{huang2015system}, \cite{hei2013two}. Huang et al.\cite{ss_attack_1} implemented a tool to find high-risk methods in System Server, and found Android Stroke Vulnerabilities by scrutinized those methods. Gu et al. \cite{Gu2017JGRE} designed a tool to examine all system services and investigated the vulnerabilities' impact. Cao et al. \cite{cao2015towards} use fuzzer to scan all system services to found input validation vulnerability. Zhang et al. \cite{invetter} developed Invetter, which can find vulnerable input validations by using machine learning and static analysis. Armando et al. \cite{armando2012would} carry out a DoS attack that forks excessive Zygote processes to make Android unresponsive and present two fixes. Huang et al. \cite{huang2015system} performed DoS attacks by exploiting the vulnerability in the concurrency control of Android system services. Hei et al. \cite{hei2013two} present two vulnerabilities in Android and one of the vulnerabilities can be exploited to launch the DoS attack.

\section{Conclusion}
In this paper, we systematically studied the attacks and defenses of JNI Global Reference resource exhaustion in the Android system. We built an attack which can lead the Android system infinitely reboot. We designed JGREAnalyzer to analyze the vulnerabilities in Android. As a result, we discover 34 JGRE vulnerabilities in Android 8.1, 28 JGRE vulnerabilities in Android 9, and 21 JGRE vulnerabilities in Android 10. Furthermore, we developed JGRE Purger to defend against all the existing JGRE attacks.

\section*{Acknowledgement}
This work was partially supported by the National Natural Science Foundation of China under grants U1736209, 61872438 and 61572278, Leading Innovative and Entrepreneur Team Introduction Program of Zhejiang (2018R01005).  


\bibliographystyle{IEEEtran}
\bibliography{main}

\balance

\vspace{-2.5em}
\begin{IEEEbiographynophoto}{Yi He} received his Master's degree from Tsinghua University. 
His current research interests include system security and program analysis. 
\end{IEEEbiographynophoto}

\begin{IEEEbiographynophoto}{Yuan Zhou} is a Master student at Tsinghua University. His current research areas are Android security and blockchain security. 
\end{IEEEbiographynophoto}

\vspace{-2.5em}

\begin{IEEEbiographynophoto}{Yajin Zhou} is currently a ZJU100 Young Professor at Zhejiang University.
He earned his Ph.D. in Computer Science from North Carolina 
State University, and then worked as a senior security researcher at
Qihoo 360. His research focuses on developing practical solutions to
improve mobile security and privacy, mainly from the aspect of
enriching the app ecosystem.
\end{IEEEbiographynophoto}

\vspace{-2.5em}

\begin{IEEEbiographynophoto}{Qi Li} received his Ph.D. degree from Tsinghua University. Now he is an associate professor at Institute for Network Sciences and Cyberspace, Tsinghua University. He has ever worked at ETH Zurich and the University of Texas at San Antonio.
His research interests are in network and system security, particularly in Internet and cloud security, mobile security, and big data security. He is currently an editorial board member of IEEE TDSC and ACM DTRAP.
\end{IEEEbiographynophoto}

\begin{IEEEbiographynophoto}{Kun Sun} is an associate professor at George Mason University. He received his Ph.D. in Computer Science from North Carolina State University. His research focuses on systems and network security. Dr. Sun has more than 15 years working experience in both industry and academia, and serves as the Chief Scientist of the Center for Secure Information Systems (CSIS) and the director of the Sun Security Laboratory (SunLab), which is continuously hiring self-motivated undergraduate and graduate students who have research interests on information security, operating system, and computer networks.
\end{IEEEbiographynophoto}

\vspace{-2.5em}

\begin{IEEEbiographynophoto}{Yacong Gu} received his Ph.D. degree from the Chinese Academy of Science in 2016. His current research interests include mobile security, in particular Android security. 
\end{IEEEbiographynophoto}

\begin{IEEEbiographynophoto}{Yong Jiang} received his Ph.D. degree from Tsinghua University. Now he is a professor at Graduate School at Shenzhen, Tsinghua University. His current research interests include computer networks and network security. 
\end{IEEEbiographynophoto}


\end{document}